\documentclass[aps,twocolumn]{revtex4-2}

\usepackage{graphicx}
\usepackage{bm}
\usepackage{amsmath}
\usepackage{amssymb}
\usepackage{euscript}
\usepackage{verbatim}
\usepackage{wrapfig}
\usepackage{setspace}
\usepackage{xcolor}
\usepackage{amsfonts}
\usepackage{lipsum}
\usepackage[caption=false]{subfig}
\usepackage{romannum}
\usepackage{makecell}

\usepackage[colorlinks=true, allcolors=black]{hyperref}

\newcommand{\Tr}{\mathrm{Tr}}

\newcommand{\ket}[1]{\left| #1 \right\rangle}
\newcommand{\bra}[1]{\left\langle #1 \right|}
\newcommand{\braket}[2]{\left\langle #1 | #2 \right\rangle}

\begin{document}
	
\title{Noisy Quantum Computation Modeled by Quantum Walk: Universality without Ancillas}

\author{Noa Feldman}
\author{Moshe Goldstein}
\affiliation{Raymond and Beverly Sackler School of Physics and Astronomy, Tel-Aviv University, 6997801 Tel Aviv, Israel}

\begin{abstract}
The universal quantum computation model based on quantum walk by Childs has opened the door for a new way of studying the limitations and advantages of quantum computation, as well as for its intermediate-term simulation. In recent years, the growing interest in noisy intermediate-scale quantum computers (NISQ) has lead to intense efforts being directed at understanding the computational advantages of open quantum systems. In this work, we extend the quantum walk model to open noisy systems in order to provide such a tool for the study of NISQ computers. Our method does not use explicit purification, and allows to ignore the environment degrees of freedom and obtain a much more efficient implementation (linear rather than exponential in the runtime), which employs no ancillas hence provides direct access to the entanglement properties of the system. In our scheme, the quantum walk amplitudes represent elements of the density matrix rather than the wavefunction of a pure state. Despite the non-trivial manifestation of the normalization requirement in this setting, we model the application of general unitary gates and nonunitary channels, with an explicit implementation protocol for channels that are commonly used in noise models.
\end{abstract}

\maketitle

%\section{Introduction}

\section{Introduction} Since the suggestion by Feynman to simulate quantum systems using controllable quantum systems (rather than classical computers)~\cite{Feynman1982}, a variety of possible realizations have been suggested. Such realizations have been shown to be advantageous in additional computational tasks over classical computers~\cite{Shor94,Grover1997}.

Several models of quantum computers have been suggested, the most common of which include the circuit~\cite{NielsenChuang}, adiabatic~\cite{Annealing94,Annealing98,Annealing00,Annealing02,AdiabaticMay03,AdiabaticMay04,AdiabaticMay06,AdiabaticMay07,Adiabatic09,Annealing20}, and measurement-based quantum computation~\cite{Briegel2009}. 
%and boson sampling~\cite{Sampling10,Sampling15}.
A model that can solve any problem that can be solved on a quantum Turing machine is called \textit{universal}, and all of the models just listed have been shown to be have this property. Studying the same problem on different models allows to examine it from different angles and may provide a better understanding of quantum advantages and limitations.

One such universal model was suggested by Childs~\cite{Childs}, and describes a single particle performing a quantum walk on a graph~\cite{FarhiGutmanAug98,WatrousDec98,AharonovEt01,Review12}. Quantum walk models have been shown to have an advantage over classical random walk, with~\cite{ChildsEt03,ChildeEt07} and without~\cite{QWMay03,QW_04,QWAug04,QW_05,QWJun05,QW_06,QW_07,QWJun07,QW_08,QW12} a black-box, a property which can be leveraged for developing quantum algorithms. The model opens the door for using well-developed tools in physics towards studying quantum computation, such as scattering and localization, allowing for numerous algorithms and results~\cite{QWAug04,search3,search4,search5,search6,search7,search8,search9,search10,search11,search12,search13}. It also allows the experimental simulation of near-term quantum circuits in various platforms~\cite{exp_Oct08,exp_Jul09,exp_Mar10,exp_Apr10,exp_Sep10,exp_Jul12,exp_Jul17,exp_Jan18,exp_Jan19,exp_May20,exp_Jul20}. This includes photonics with classical light, which can serve as an intermediate step in the development of the controllable photonic hardware needed for linear optical quantum computation~\cite{optics07,optics21,optics21Bartolucci}.

As explained in detail in Sec. \ref{childs}, 
the quantum walk computational model uses a particle (walker) scattered on a graph, where each node in the graph corresponds to a wavefunction amplitude. The particle's wavefunction can simulate the wavefunction of a quantum register during an application of a noiseless quantum circuit. The straightforward way of adding noise to the simulated system is purification, that is, adding ancillary qubits to the  circuit and using them as an environment which interacts with the original circuit. This method treats the simulated ``environment'' and ``system'' on an equal footing, and makes it harder to study the effects of noise on the quantum computer itself. Additionally, the purification process is costly: Each interaction with the environment requires an additional environment qubit, thus doubling the graph size and making the complexity exponential in the runtime.

In recent years, the interest in the study of noisy quantum systems is growing~\cite{Noisy98,Noisy2018,Noisy2019,Noisy2021}, as its understanding is a key component in characterizing the computational strength of noisy intermediate-scale quantum computers (NISQ)~\cite{NISQ}. This calls for an extension of the quantum walk model that accounts for the effects of noise without modeling the environment (so as to keep the cost linear in the runtime), which is the goal of this present work. For an open system described by a density matrix (DM) $\rho$, we perform a vectorization of $\rho$ and treat the DM elements as the basic degrees of freedom in the model, to which individual nodes in the graph correspond. Naively one would expect such an approach to face difficulties: In a quantum walk, the sum of the absolute values of the amplitudes square is preserved, while for a DM we should instead demand the preservation of the sum of elements on the diagonal ($\Tr(\rho)$). However, the new requirement is surprisingly obeyed for a significant set of operators, and can be handled fully by appropriate modifications of the rest.
Unitary operations on the DM turn out to be implemented using unitary operations on its elements, as if they were the amplitudes of a pure state. Nonunitary channels are described using additional outgoing wires, but without explicit purification or ancillas. The entanglement between the noisy system and its environment, which is an important indicator for the performance of the quantum computer, can be naturally extracted from the quantum walker's state.

The paper is organized as follows: In Sec. \ref{childs} we review the pure state model introduced by Childs ~\cite{Childs}. In Sec. \ref{sec:open} we describe our extension of this model to mixed states. We discuss the representation of DMs and their entanglement in the model, and describe an implementation of a universal set of unitaries applied to DMs. In Sec. \ref{sec:nu} we develop the realization of nonunitary channels in the system, with examples of the depolarizing channel in Sec. \ref{sec:exampledepol} and the erasure channel in Sec. \ref{sec:exampleerase}. Appendix \ref{appendix} provides an example for the calculation of the unitary implemented by a widget, while Appendix B analyzes the graph size in our approach as opposed to employing purification.

%\section{The Childs Quantum Walk Model}\label{sec:pure}
\section{Childs' pure state quantum walk model}\label{childs}
\begin{figure}[t]
{		
	\includegraphics[width=\linewidth]{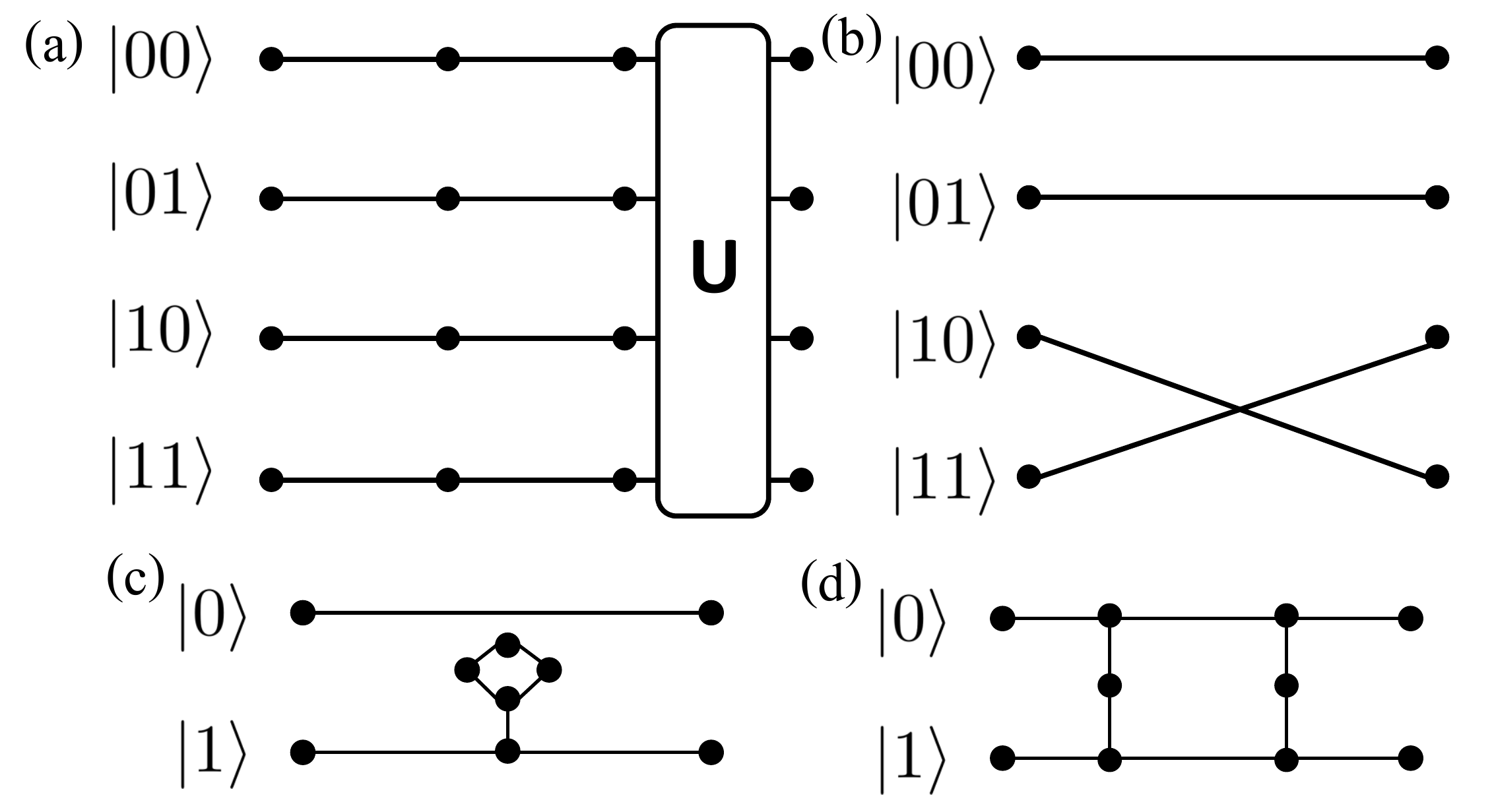} 
} 
\caption{(a) Example of a quantum walk graph for a pure state of two qubits over three timesteps, where evolution in time is represented by moving right on a wire in the graph. In the first two timesteps, no unitary is applied. In the third timestep, a unitary $U$ is applied. The unitaries are implemented by graph widgets constructed of additional nodes and edges (designed to guarantee zero backscattering for $k = \pi/4$), such as the CNOT gate demonstrated in (b). A particle entering the graph at the top left node simulates a quantum computation starting at $\ket{00}$. (b) Realization of the CNOT gate for two qubits, where the control qubit is the left one. An application of a CNOT gate does not change the amplitudes of $\ket{00}, \ket{01}$ and switches the amplitudes of $\ket{10}, \ket{11}$. (c) Implementation of the one-qubit phase gate $U_1$ defined in Eq.~(\ref{eq:us}). A particle with momentum $k=\frac{\pi}{4}$ acquires a phase of $\frac{\pi}{4}$ going through the widget added to the $\ket{1}$ wire (see Appendix \ref{appendix}). (d) Implementation of the one-qubit gate $U_2$ defined in Eq.~(\ref{eq:us}). }
\label{fig:pureWidgets}
\end{figure}

We now briefly review Childs' quantum walk model for pure state quantum computation~\cite{Childs}. Each state in the Hilbert space of the $n$ qubits, in each timestep, is represented by a node in a graph, as depicted in Fig.~\ref{fig:pureWidgets}(a). 
The particle is controlled by a tight-binding Hamiltonian identical to the adjacency matrix of the graph.
We start by sending a single particle (walker) wavepacket into the graph from the left, on the wire corresponding to $\ket{0}^{\otimes n}$. We filter the particle's momentum to be $k = \pi/4$. We then let the particle evolve under the graph Hamiltonian, i.e., be scattered through the graph. Quantum gates are implemented by graph widgets, designed such that no backscattering occurs, only forward scattering. Hence, the widgets naturally preserve the sum of amplitudes squared in each timestep, and automatically map onto unitary operations in the qubit Hilbert space. The simplest example of such a graph widget is the one implementing the CNOT gate, depicted in Fig.~\ref{fig:pureWidgets}(b). A universal set of gates can be obtained by adding the following one-qubit unitaries to the CNOT gate:
\begin{equation}\label{eq:us}
	\begin{aligned}
	 U_1 = \begin{pmatrix}
		1 & 0 \\ 0 & e^{i\pi/4}
	\end{pmatrix}, U_2 = \frac{1}{\sqrt{2}}\begin{pmatrix}	i & 1 \\ 1 & i \end{pmatrix}.
	\end{aligned}
\end{equation} 
The widgets implementing $U_1$ and $U_2$ are proposed by Childs~\cite{Childs} and reproduced in Figs.~\ref{fig:pureWidgets}(c) and \ref{fig:pureWidgets}(d), respectively.
An explicit calculation is given in Appendix \ref{appendix} for the widget implementing $U_2$. For the chosen momentum $k=\pi/4$, the reflection coefficient from all widgets is 0, and the transmission coefficients into different wires correspond to the unitary implemented by the widget; these properties are insensitive to small variations in $k$~\cite{Childs}.

The space complexity of the graph is linear in the Hilbert space size, which is exponential in the number of qubits $n$, and linear in the number of gates in the simulated quantum circuit, that is, the runtime of the quantum algorithm. Thus, although this approach is impractical for the experimental realization of large systems, the fact that it only involves linear elements still gives it advantage in the near future~\cite{exp_Oct08,exp_Jul09,exp_Mar10,exp_Apr10,exp_Sep10,exp_Jul12,exp_Jul17,exp_Jan18,exp_Jan19,exp_May20,exp_Jul20}, especially in optical systems, where strong nonlinearities are hard to achieve. More importantly, the description of the quantum circuit model as a particle scattered on a graph allows analyzing its behavior in the terms of scattering phenomena, such as diffusion or Anderson localization~\cite{banff,Akkerman}, and thus to establish its quantum advantage~\cite{gluedTrees}  even for so-called ``stoquastic'' Hamiltonians~\cite{stoq1,stoq2,stoq3}. 

%\section{Open systems}\label{sec:open}
\section{Open Systems: Representing the Density Matrix and Unitary Gates}\label{sec:open}
A noisy quantum system interacts with its environment, and as such it can be modeled as the original system and additional ancillary qubits that account for the environment. The system and the new ``environment'' are treated as a system in a pure state, and the process is referred to as \textit{purification}. Such a purification can in principle be implemented in the quantum walk model, by introducing additional qubits (ancillas) to the system, which will be reflected by an increase in the size of the Hilbert space and thus an increase in the number of wires. However, since each wire represents a state in the the full system's Hilbert space (rather than, say, a single qubit), each interaction of a system qubit with the environment requires doubling the amount of wires in the circuit, leading to an exponential scaling of the cost with the runtime. Moreover, the purified system cannot be separated into the studied system and the purifying qubits. When studying noisy quantum computers, the state of the environment is unknown and unimportant, making such purification very inconvenient. For these reasons we generalize the original model into a model that separates the studied system from its environment and allows to ignore the environment state. Furthermore, the graph size is linear in the runtime, as we further discuss below in Sec. \ref{sec:nu} and in Appendix \ref{appendix:graph_size}.

We extend the method from pure states to DMs by assigning a single wire to each DM element, which sums up to $2^{2n}$ wires for a system of $n$ qubits. As mentioned above, the extension is not straightforward, since scattering on a graph allows for unitary operators on a particle scattering forward, which are equivalent to the unitary evolution of a state vector, $U\ket{\psi}$, as opposed to the unitary evolution of a DM, $U\rho U^\dagger$; equivalently, the walk preserves the sum of the absolute values squared of the amplitudes, but we would like to preserve $\mathrm{Tr}(\rho)$, which is a sum of some of the amplitudes (those corresponding to elements on the diagonal). Nevertheless, as we now show, implementing unitary operations on our DM representation is possible within the quantum walk toolkit. Moreover, the entanglement of the system nicely emerges from the walker's probability density.

\begin{figure}[t]
	{		
		\includegraphics[width=\linewidth]{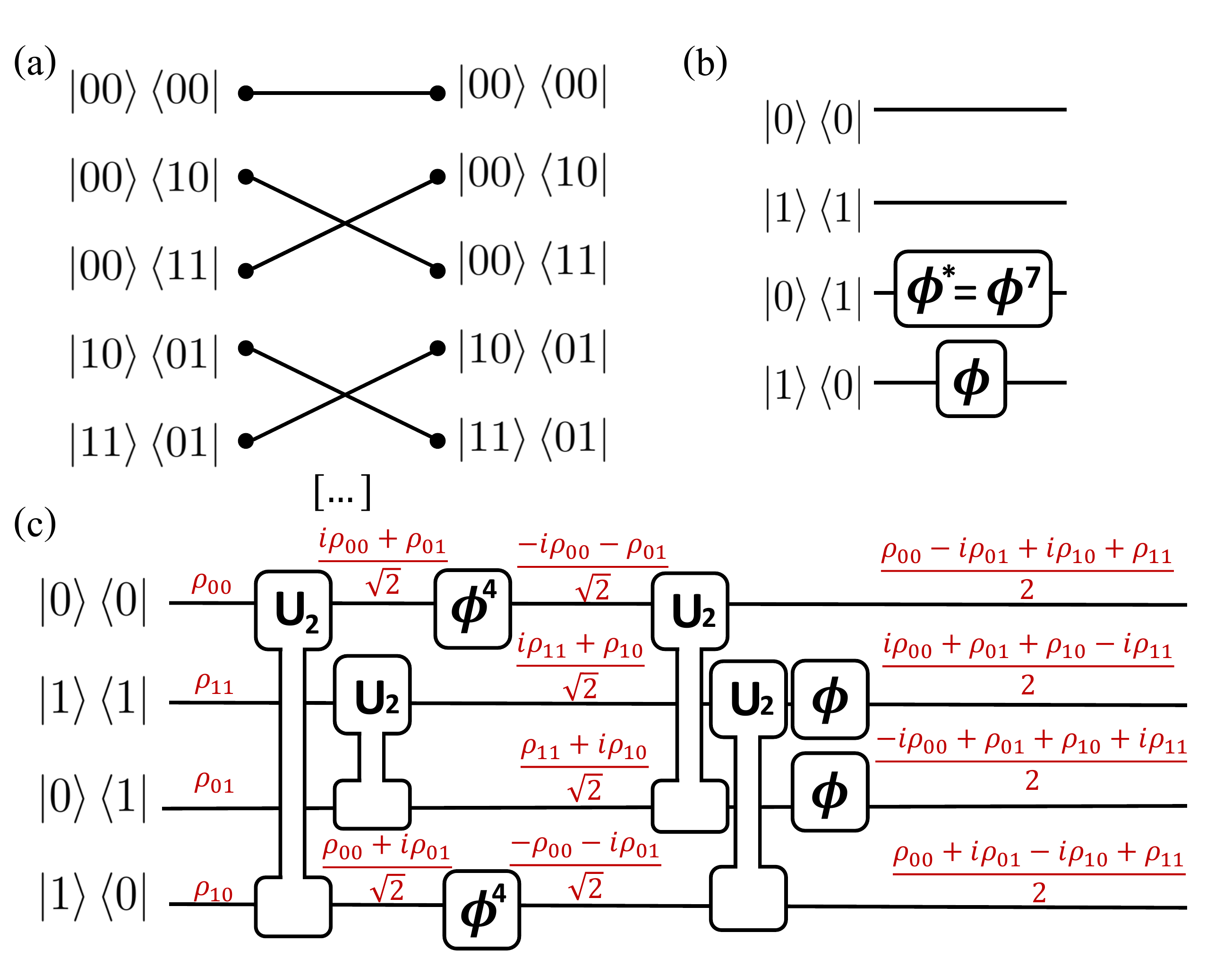} 
	} 
	\caption{Widgets of the universal gate set $\{\text{CNOT}, U_1,U_2\}$ for a DM,
	i.e., the implementation of $\rho \to U\rho U^\dagger$ for each unitary $U$ in the set.  (a) Implementation of CNOT acting on a DM, where the control qubit is the left qubit. For simplicity, only a subset of the wires is shown. (b) Application of the gate $U_1$ defined in Eq.~(\ref{eq:us}), Here the rectangles represent an element that adds a phase $\phi=e^{i\pi/4}$ which is implemented as in Fig.~\ref{fig:pureWidgets}(c), as well as $\phi^*=\phi^7$. (c) Application of the gate $U_2$ defined in Eq.~(\ref{eq:us}), using two instances of Child's 2-wire $U_2$ widgets depicted in Fig.~\ref{fig:pureWidgets}(d). The red expressions give the evolution of the amplitudes along the circuit. }
	\label{fig:mywidgets}
\end{figure}

We start by addressing the implementation of a universal set of unitary gates. We use the same widgets suggested by Childs to filter the momentum of the walkers to $k=\pi/4$. The open system widgets therefore function similarly to the pure state case. We rely on the pure state gate set designed by Childs: the two-qubit CNOT gate and the one-qubit gates $U_1$ and $U_2$ defined in Eq.~(\ref{eq:us}). This implementation is presented in Fig.~\ref{fig:mywidgets}. Sending a particle with momentum $k=\pi/4$ through the node $\ket{\text{start}}\bra{\text{start}}$ at $t=0$ corresponds to starting the quantum computation when the qubits are at the pure state $\ket{\text{start}}$. After scattering through the graph, the amplitudes on the wires are equal to the desired elements of the DM after the application of the quantum computation.

The trace of $\rho$ is conserved throughout all gate applications. The particle density amplitude on the outgoing wires to the right of the graph is the value of the corresponding density elements. The walker's probability density on the wire corresponding to $\rho_{ij}$ is  $|\rho_{ij}|^2$. This is the contribution of this element to the purity, which is defined by
\begin{equation}\label{eq:purity}
p_2 = \Tr\left(\rho^2\right) = \sum_{i,j}\rho_{ij}\rho_{ji}. 
\end{equation}
The purity is the second R\'enyi moment, and is often used as an entanglement measure~\cite{Wilde}: the smaller the purity, the more entangled the state is.
Summing the walker's probability density over all the wires directly gives the purity; the more the density function is spread across the wires, the larger the entanglement.
Estimating the purity can become useful when applying nonunitary gates to the initially pure DM we start from, as described in Sec. \ref{sec:open}.
%as described in section \ref{sec:depol}. 

The purity can also be estimated for some subsystem $A$ of the simulated system, by performing a partial trace over its complement $\bar{A}$, $\rho_A = \Tr_{\bar{A}}(\rho)$. Applying a partial trace over a single qubit on the graph 
is shown in Fig.~\ref{fig:examples}(a). The Hadamard gate, 
\begin{equation}\label{eq:hadamard}
	H = \frac{1}{\sqrt {2}}\begin{pmatrix}	1 & 1 \\ 1 & -1 \end{pmatrix} = U_1^2U_2U_1^2,
\end{equation}
is applied to all pairs of wires corresponding to elements of the form $\ket{\alpha 0}\bra{\beta 0}, \ket{\alpha 1}\bra{\beta 1}$, where the qubit being traced out is the rightmost qubit. The first outgoing wire carries an amplitude of $(\rho_{\alpha 0,\beta 0}+\rho_{\alpha 1,\beta 1})/\sqrt{2}$, which, up to a factor of ${1}/{\sqrt{2}}$, equals the $\alpha,\beta$ element of the traced out DM. The second outgoing wire carries an amplitude of $(\rho_{\alpha 0,\beta 0}-\rho_{\alpha 1,\beta 1})/\sqrt{2}$, which is discarded, along with the wires representing $\rho_{\alpha 0,\beta 1}$, $\rho_{\alpha 1,\beta 0}$. Summing the walkers' probability density over the kept wires therefore gives the purity of A times 1/2. One can trace out additional system qubits similarly.

%\section{Applying nonunitary Operators}\label{sec:depol}
\section{Applying nonunitary channels}\label{sec:nu}
Open systems which interact with an environment undergo nonunitary channels on top of the unitary gates. We now review the model of such channels and develop a method for implementing them.

A nonunitary channel can be described as 
\begin{equation}
	\rho \rightarrow \sum_i \mathcal{K}_i \rho \mathcal{K}_i^\dagger,
\end{equation}
using the so-called Kraus operators $\mathcal{K}_i,$~\cite{BP} which obey the normalization requirement
	$\sum_i \mathcal{K}_i^\dagger \mathcal{K}_i = \mathbb{I}.$

An interesting channel often used to model interaction of a quantum computer with an environment is the depolarizing channel with probability $p$:
\begin{equation}\label{eq:depol}
	\mathcal{K}_{1\dots4} = \sqrt{1 - p}\mathbb{I}, \sqrt{\frac{p}{3}}\sigma^x, \sqrt{\frac{p}{3}}\sigma^y, \sqrt{\frac{p}{3}}\sigma^z, 
\end{equation}
where $\sigma^i$ is the $i$th Pauli matrix. The depolarizing channel models a spontaneous measurement of the qubit by the environment, with the measurement outcome unknown. 

%Some additional interesting channels are the phase damping channel:
%\begin{equation}\label{eq:phaseDamping}
%	\mathcal{K}_{1\dots3} =  \sqrt{1 - p}\mathbb{I}, \sqrt{p}\ket{0}\bra{0}, \sqrt{p}\ket{1}\bra{1}, 
%\end{equation}
An additional interesting channel is the erasure channel, 
\begin{equation}\label{eq:erasure}
	\rho \rightarrow \ket{0}\bra{0}, \quad  \mathcal{K}_{1,2} = \ket{0}\bra{0}, \ket{0}\bra{1},
\end{equation}
which models a qubit loss (especially important in photon-based systems) and its replacement by a new qubit initiated to $\ket{0}$.

For simplicity, we concentrate below on 1-qubit channels. Such channels, specifically the channels demonstrated above, are used in many of the noise models in NISQ computers. The following analysis can be generalized to many-qubit channels. 

In general, the implementation of nonunitary channels requires the addition of ancilla wires (though much fewer than purification). To see that, let us note that in a scattering problem, we can only apply unitary operators to the wires. Combined with the requirement that the channel is trace-preserving, we see that applicable channels always conserve the polarization of the single-qubit DM. This becomes clear when writing $\rho = \left(a_0\mathbb{I} + a_1X + a_2Y  +a_3Z\right)/2$, which in fact can be achieved in our DM walk model by a unitary transformation of the amplitudes, $\begin{pmatrix} \rho_{00} & \rho_{01} & \rho_{10} & \rho_{11} \end{pmatrix}^T \rightarrow \begin{pmatrix}
a_0 & a_1 & a_2 & a_3 
\end{pmatrix}^T/2$. The trace-conservation requirement now translates into $a_0=\text{const.}$ Any unitary applied can then only mix the coefficients of $X, Y, Z$, conserving the polarization $\langle X \rangle^2 + \langle Y \rangle^2 + \langle Z \rangle^2$. The relation between polarization and entanglement becomes clear in this interpretation: Such a unitary does not change the purity of the qubit under consideration (as defined in Eq.~(\ref{eq:purity})), that is directly related to the amplitude on the wires. Therefore, to implement nonunitary operations which change the polarization we need additional wires, as we explain below.

Any single 1-qubit channel can be expressed as a $4\times4$ matrix $O$ applied to the vectorized DM. We can perform a singular value decomposition (SVD)~\cite{Schollwock} of this matrix, $O = USV^\dagger$, where $U,V$ are unitary and $S$ is diagonal with non-negative real elements, the singular values of $O$. Being unitary, $U, V$ can always be implemented using Childs' widgets. Therefore, the nontrivial step is the multiplication by $S$.

We separate the channels into two groups: channels that never increase the purity and polarization (i.e., never decrease the entanglement of the system with its environment), which correspond to unital maps, such as the depolarizing channel (\ref{eq:depol}), and channels that can increase the purity, which correspond to nonunital maps, such as the erasure channel (\ref{eq:erasure}). 
% and the phase-damping channel (\ref{eq:phaseDamping}),

In the first case, the application of the channel can only decrease the amplitude on the graphs. We are thus guaranteed that the singular values are $\le 1$. The application of S can be done by damping amplitude out of the graph; since $S$ is diagonal, this can be implemented by adding a half-infinite wire that ``leaves'' the graph to each wire in the graph. We note that these additional wires are added only where nonunitary channels are applied, as opposed to purification, where the doubling of the number of wires due to each ancilla qubit needs to be followed along the entire graph. We denote by $D(\lambda)$ the amplitude damping widget, which realizes the isometry
\begin{equation}\label{eq:dlambda}
	D(\lambda) = \begin{pmatrix}
		\lambda & \sqrt{1-\lambda^2}
	\end{pmatrix}^T,
\end{equation}
where the top and bottom elements correspond to the wires staying and leaving the graph, respectively.
One possible implementation for $D(\lambda)$ is illustrated in Fig.~\ref{fig:examples}(b). $D(\lambda)$ depends on the singular value $\lambda$ and therefore it includes $\lambda$-dependent hopping parameters on its edges:
\begin{equation}\label{eq:dparams}
	\begin{aligned}
		j_1^\lambda = \lambda, \quad
		j_2^\lambda = \sqrt{1-\lambda^2}.
	\end{aligned}
\end{equation}
We note that the widget $D(\lambda)$ is completely insensitive to the value of $k$. The implementation above requires a small number of nodes and edges, at the cost of changing the hopping amplitudes locally. A second possible implementation would be to use the universal set of single-qubit unitaries $U_1, U_2$ to implement the unitary:
\begin{equation}\label{eq:Ud}
	U_D(\lambda) = \begin{pmatrix}
		\lambda & -\sqrt{1-\lambda^2} \\ \sqrt{1-\lambda^2} & \lambda
	\end{pmatrix}.
\end{equation}
The widget can be applied to a single ingoing wire, which corresponds to the ingoing vector being in the state $\begin{pmatrix}
	1 & 0
\end{pmatrix}^T$. In this way, a more complicated set of widgets is required (for most $\lambda$s), but the hopping parameter on all edges remains constant.
A full implementation of the depolarizing channel is given in Sec. \ref{sec:exampledepol} below.

In the second case, where the purity may grow, we need to increase the amplitudes on the wires. In order to do this without keeping track of the state of the environment (from which the amplitude increase emerges), the channel is implemented by rescaling of the amplitudes on the graph: We rescale $S \rightarrow \frac{1}{\lambda_{\mathrm{max}}}S$, where $\lambda_{\mathrm{max}}>1$ is the largest singular value of the channel. We then apply the amplitude damping widget $D(\lambda)$ to the wires not corresponding to $\lambda_{\mathrm{max}}$. Postprocessing the amplitudes on the graph is done with consideration of this rescaling by the (known) $\lambda_{\mathrm{max}}$. An implementation of the erasure channel is presented in Sec. \ref{sec:exampleerase} below.

A comparison of the required graph size for dissipative systems in both models --- pure state with purification and the open case suggested in this work --- is provided in Appendix \ref{appendix:graph_size}. It is shown that our scheme replaces the exponential dependence of the resources on the runtime in the purification approach by a linear dependence (at the lesser cost of stronger exponential dependence on the number of simulated qubits).
Let us also add that whether a rescaling of the amplitudes is done or not, damping amplitude out of the graph results in a possibility to measure the quantum walker outside of the graph. This property is an expression of the possibility to get a false result (or detect an error) in a noisy quantum computer. A bound for the survival probability is provided in Appendix \ref{appendix:graph_size}, showing that even when taking it into account, our approach keeps its advantage over using ancillas.  We further note that the probability that the walker remains in the graph equals the resulting state's purity, Eq.~(\ref{eq:purity}); this direct access to the purity is another merit of our approach.

%\subsubsection{The Depolarizing Channel}\label{sec:depolexample}
\subsection{Example 1: The depolarizing channel}\label{sec:exampledepol}
\begin{figure}[t!]
	{		
		\includegraphics[width=\linewidth]{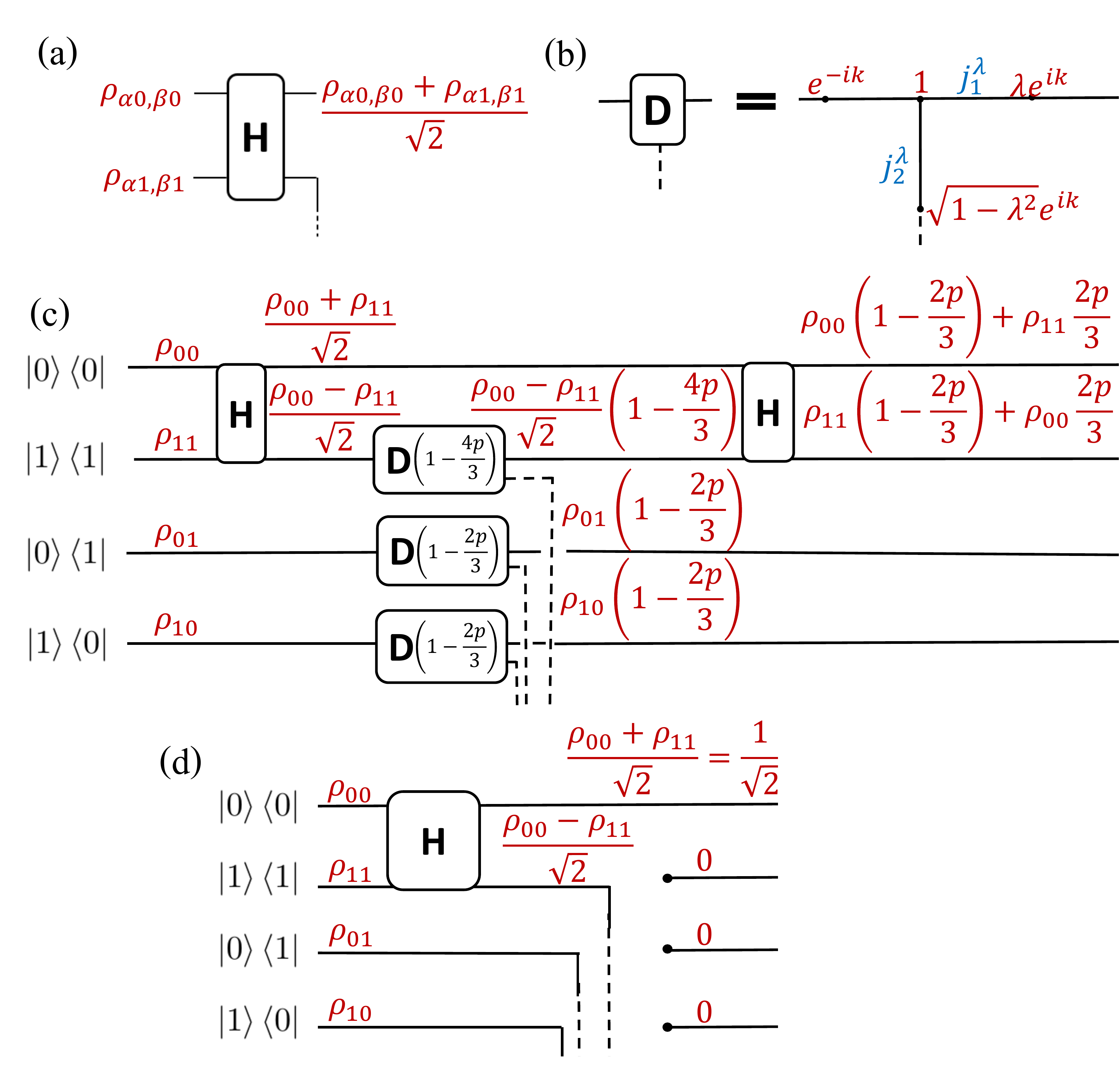} 
	}
	\caption{(a) Implementation of a partial trace on the right qubit. (b) The graph widget $D(\lambda)$. In blue are $\lambda$-dependent hopping parameters defined in Eq.~(\ref{eq:dparams}), and in red are the amplitudes on the nodes for an ingoing particle from the left. One may avoid the use of $\lambda$-dependent hopping parameters by using a (generally more complicated) combination of the widgets for $U_1$ and $U_2$ (cf. Eq.~(\ref{eq:Ud})). (c-d) The graph widgets implementing (c) the depolarizing channel and (d) the erasure channel, based on their SVD, see main text.}
	\label{fig:examples}
\end{figure}

We write the depolarizing channel as the matrix that acts on the wires and apply SVD to it:
\begin{equation}
\begin{aligned}
&\begin{pmatrix}
\rho_{00}  \\ \rho_{11} \\ \rho_{01} \\ \rho_{10}
\end{pmatrix} \rightarrow 
\begin{pmatrix}
1 - \frac{2p}{3} & \frac {2p}{3} & 0 & 0 \\
\frac {2p}{3} & 1 - \frac{2p}{3} & 0 & 0 \\
0 & 0 & 1 - \frac{2p}{3} & 0 \\
0 & 0 & 0 & 1 - \frac{2p}{3}
\end{pmatrix}\begin{pmatrix}
\rho_{00}  \\ \rho_{11} \\ \rho_{01} \\ \rho_{10}
\end{pmatrix} = \\&
U
\begin{pmatrix}
1 & 0 & 0 & 0 \\ 0 & 1 - \frac{4p}{3} & 0 & 0 \\ 0 & 0 & 1-\frac{2p}{3} & 0\\ 0 & 0 & 0 & 1-\frac{2p}{3}
\end{pmatrix}
U^\dagger
\begin{pmatrix}
\rho_{00}  \\ \rho_{11} \\ \rho_{01} \\ \rho_{10}
\end{pmatrix},
\end{aligned}
\end{equation}
where, in $2\times 2$ block notation,
\begin{equation}\label{eq:svdU}
U = \begin{pmatrix}
H & 0 \\ 0 & \mathbb{I}_2
\end{pmatrix},
\end{equation}
which can be implemented be applying a Hadamard gate defined in Eq.~(\ref{eq:hadamard})
%(as discussed at the end of Sec. \ref{sec:open})
to the wires corresponding to $\rho_{00},\rho_{11}$. The full widget combination implementing the channel is depicted in Fig.~\ref{fig:examples}(c).

%\subsubsection{The Erasure Channel}
\subsection{Example 2: The erasure channel}\label{sec:exampleerase}
%The erasure channel is very similar to the partial trace 
%demonstrated at the end of Sec. \ref{sec:open}.
The erasure channel is similar to the partial trace 
demonstrated above.
This is expected, as an erasure channel describes ignoring a qubit (due to loss, for example) and replacing it with a new qubit initiated to 0, which is equivalent to tracing that qubit out. The matrix implementing the channel and its SVD is:
\begin{equation}
\begin{aligned}
&\begin{pmatrix}
\rho_{00}  \\ \rho_{11} \\ \rho_{01} \\ \rho_{10}
\end{pmatrix} \rightarrow 
\begin{pmatrix}
1 & 1 & 0 & 0 \\
0 & 0 & 0 & 0 \\
0 & 0 & 0 & 0 \\
0 & 0 & 0 & 0
\end{pmatrix}\begin{pmatrix}
\rho_{00}  \\ \rho_{11} \\ \rho_{01} \\ \rho_{10}
\end{pmatrix} = \\&
\mathbb{I}
\begin{pmatrix}
\sqrt{2} & 0 & 0 & 0 \\ 0 & 0 & 0 & 0 \\ 0 & 0 & 0 & 0 \\ 0 & 0 & 0 & 0
\end{pmatrix}
U^\dagger
\begin{pmatrix}
\rho_{00}  \\ \rho_{11} \\ \rho_{01} \\ \rho_{10}
\end{pmatrix},
\end{aligned}
\end{equation}
where $U$ is defined in Eq.~(\ref{eq:svdU}). We can apply $U$ to the wires, damp the amplitude on the wires corresponding to singular values 0, and rescale the amplitudes on the graph by $\sqrt{2}$. The full widget combination implementing all stages but the rescaling (which is deferred to postprocessing) is depicted in Fig.~\ref{fig:examples}(d).

%\section{Conclusions}
\section{Conclusion}
The quantum walk computation model by Childs~\cite{Childs} is universal, and as such is useful for the study of quantum computation advantages~\cite{ChildsEt03,ChildeEt07, QWMay03,QW_04,QWAug04,QW_05,QWJun05,QW_06,QW_07,QWJun07,QW_08,QW12,search3,search4,search5,search6,search7,search8,search9,search10,search11,search12,search13}\hypersetup{citecolor=black}. With the growing interest in noisy quantum systems, an extension of the model to noisy open systems becomes desirable.

We have extended the model to treat open systems, in a way that is much more efficient than purification. The degrees of freedom of our construction represent the system alone. The purity arises naturally in the model, allowing for the analysis of the system's entanglement with its environment. Commonly used channels for modeling noise are demonstrated explicitly, and a general protocol is provided for all other possible channels.

The model can now be used for the analysis of entanglement in quantum noisy channels and for studying the effect of random unitaries on a quantum system coupled to an environment~\cite{percolation,Ruhman17,LiFisher_Nov18,LiFisher_Oct19,VasseurLudwig_Oct19,Chan_Jun19,Skinner_Jul19,BaoAltman20,ChoiAltmanJul20}. The corresponding graph widgets used for simulating noise can be analyzed using tools from, e.g., Anderson localization theory~\cite{banff,Akkerman} and may provide new insights on the computational power of quantum noisy circuits. Experimental realizations using either quantum particles or classical light are possible for intermediate-size systems.

We thank D. Aharonov for stimulating discussions, in which the problem considered in this work came up.
Our work has been supported by the Israel Science Foundation (ISF) and the Directorate for Defense Research and Development (DDR\&D) grant No. 3427/21 and by the US-Israel Binational Science Foundation (BSF) Grants No. 2016224 and 2020072.
NF is supported by the Azrieli Foundation Fellows program.

\appendix
\section{Explicit derivation of widget functionality}\label{appendix}
\begin{figure}[t]
		{		
			\includegraphics[width=0.7\linewidth]{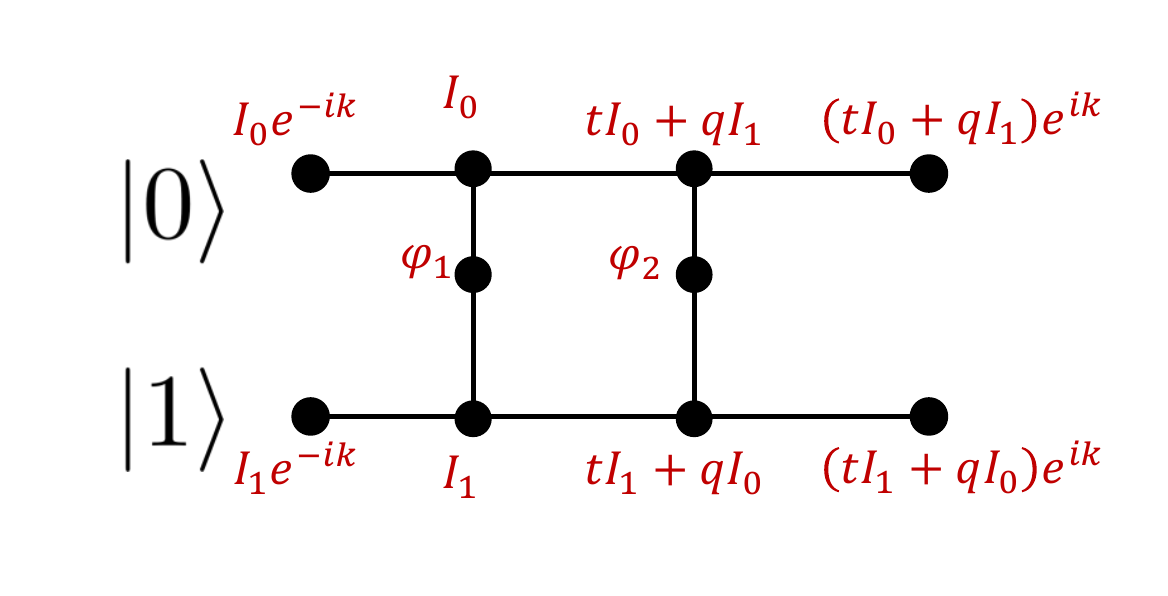} 
		}
		\caption{The widget implementing $U_2$ from Eq.~(\ref{eq:us})  and presented in Fig.~\ref{fig:pureWidgets}(d) with the wavefunction amplitudes annotated.}
		\label{fig:appendix}
	\end{figure}

	Here we show how the unitary implemented by a given widget can be found, using as an example the widget designed by Childs for $U_2$ and defined in Eq.~(\ref{eq:us}). The widget is depicted in Fig.~\ref{fig:appendix}; the nodes are annotated by the amplitudes of a (non-normalized) state, resulting from the the incoming state $I_0\ket{0} + I_1\ket{1}$. The choice of the amplitudes requires zero reflection by the widget, and the matrix applied to the state is 
	\[ U = \begin{pmatrix}
		t & q \\ q & t
	\end{pmatrix}. \]
	The set of equations which determine $t$ and $q$ are the Schr\"odinger equations $\bra{j}H\ket{\psi} = E\braket{j}{\psi}$ for each node $j$ of the widget,
	\begin{equation}
		\begin{aligned}
			I_0e^{-ik}+\varphi_1 + tI_0 + qI_1 &= 2\cos k I_0, 
			\\ I_0 + I_1 &= 2\cos k \varphi_1,
			\\ I_1e^{-ik} + \varphi_1 + tI_1 +qI_0 &= 2\cos k I_1,
			\\ I_0 +\varphi_2 + (tI_0 + qI_1)e^{ik} &= 2\cos k (tI_0 + qI_1),
			\\ tI_0 + qI_1 + tI_1 + qI_0 &= 2\cos k \varphi_2, 
			\\
			I_1 + \varphi_2 + (tI_1 + qI_0)e^{ik} &= 2\cos k(tI_1 + qI_0).
		\end{aligned}
	\end{equation}
	The solution of the sets of equations above results in $t=\frac{i}{\sqrt{2}}, q = \frac{1}{\sqrt{2}}$, as in Eq.~\ref{eq:us}.
	
\section{Graph size comparison}\label{appendix:graph_size}
We now compare the required graph size for dissipative systems in both models --- pure state with purification and the open case suggested in this work. A graph simulating a circuit without dissipation with $n$ qubits and depth $T$ (which equals the number of gates) requires $2^n$ wires and $O(2^n T)$ nodes, since each gate is implemented separately on each pair or quartet of wires~\cite{Childs}. We now add a dissipation modeled by the application of nonunitary channels in a fraction $f$ of the unitary gate applications. In the pure state model with added purification, $fT$ qubits need to be added (one for each channel application), resulting in $2^{n+fT}$ wires and $O(2^{n+fT} T(1+f))$ nodes. In the open system model we suggested, $O(2^{2n} fT)$ half-infinite wires are added to the $2^{2n}$ wires corresponding to the DM elements, and the full graph is composed of $2^{2n}(1 + fT)$ wires and $O\left(2^{2n}T(1+f)\right)$ nodes.
We stress that all these $O(\cdot)$ expressions are actually accurate up to constant prefactors, which are somewhat dependent on the specific gates and channels, but are comparable for the purification and for our approach. Thus, the scaling on the number of qubits in the analysis above also applies for a moderate number of qubits, such as those relevant for NISQ applications. The increasing dependence on $n$ is compensated by the linear dependence on $fT$ whenever $fT>n$, as compared to purification. 
We note that $T$ is typically linear in $n$ even for algorithms implemented by constant depth circuits (and thus relevant to the NISQ era), as the gates are applied in parallel to all qubits. The famous quantum algorithms (which are mostly relevant to the post-NISQ era) have stronger $T(n)$ dependence; for example, in Shor's factoring algorithm~\cite{Shor94},  $T$ scales more strongly than $n^2$;  and in the HHL algorithm for solving sparse linear systems~\cite{HHL,HHLComplexity}, $T=O(n\kappa\log^3 \kappa)$, where $\kappa$ is the condition number of the matrix, which is usually polynomial in $n$.
	%	(We note that both algorithms mentioned above are not expected to be implemented on NISQ computers and are presented only as an example for the scaling of $T$ with $n$).
The regime $fT > n$ is thus readily approached already for moderate $n$ and finite $f$. The analysis above is summarized in Table \ref{table:complexity}.

\begin{table*}
	\centering
	\begin{tabular}{|c|c|c|c|c|} 
		\hline
		&\multicolumn{2}{c|} {\# Wires} &\multicolumn{2}{c|} {\# Nodes} \\
		\hline
		& Noiseless & Noisy & Noiseless & Noisy  \\
		\hline
		& & & & \\[-0.8em]
		\makecell{Purification\\ } & $2^n$ & $2^{n+fT}$ & $O\left(2^{n}T\right)$ & $O\left(2^{n+fT}T(1+f)\right)$ \\
		\hline
		& & & & \\[-0.8em]
		\makecell{Open system walk\\ } & $2^{2n}$ & $2^{2n}(1+fT)$ & $O\left(2^{2n}T\right)$ & $O\left(2^{2n}T(1+f)\right)$ \\
		\hline
	\end{tabular}
	\caption{Summary of the required graph sizes for the naive purification compared to our open system quantum walk model, for a  system of $n$ qubits on which $T$ unitary gates and $fT$ nonunitary channels are applied. While in the noiseless case, the open system model suffers from requiring to square the number of wires, when introducing nonunitary channels at a rate of $f$ per unitary gate, the open system model becomes advantageous for the common case $fT > n$. The $O(\cdot)$ expressions are accurate up to a constant factor depending on the specific gates and channels, which is comparable for the purification and for our open system approach.}
	\label{table:complexity}
\end{table*}

As noted in the main text, in our implementation the walker may leave the graph and thus not detected. As explained there, this is a feature rather than a bug when using our construction for the theoretical analysis of NISQ computation, as it allows direct access to the purity of the resulting state. When implementing our scheme in hardware, this incurs some cost --- either having to use a greater incoming power using classical light, or repetitions of the measurement if one employs a quantum walker (e.g., a single photon) and the walker is lost. Since the survival probability is the purity, it cannot decrease by more than a factor of 1/4 per nonunitary gate or $(1/2)^{2n}$ overall, hence the number of repetitions is bounded by $O(2^{2\mathrm{min}(fT,n)})$. It should be noted that even this very loose bound does not affect the scaling of the overall cost with $T$ if $fT>n$, hence does not substantially affect the advantage of our approach over using purification and ancillas.

%\bibliographystyle{myapsrev4-2}
%\bibliography{childsPaper}

%

\end{document}